\documentclass{PoS}

\title{{\underline {\bf MIMAC}}: A micro-tpc matrix for directional detection of dark matter}

\ShortTitle{MIMAC: Directional detection}

\author{\speaker{D.~Santos}, J.~Billard,  G.~Bosson, O.~Bourrion, C.~Grignon, O.~Guillaudin, F.~Mayet, J.P.~Richer\\
        LPSC, Universite Joseph Fourier Grenoble 1, CNRS/IN2P3, Institut Polytechnique de Grenoble\\
        E-mail: \email{Daniel.Santos@lpsc.in2p3.fr}}

\author{E.~Ferrer, I.~Giomataris, F.J.~Iguaz, J.P.~Mols\\
	IRFU,CEA Saclay, 91191 Gif-sur-Yvette cedex}

\author{A.~Allaoua, C.~Golabek, L.~Lebreton\\
	LMDN, IRSN Cadarache, 13115 Saint-Paul-Lez-Durance} 

\abstract{Directional detection of non-baryonic Dark Matter is a promising search strategy for discriminating  WIMP events from background. However, this strategy requires both a precise measurement of the energy down to a few keV and 3D reconstruction of tracks down to a few mm. To achieve this goal, the MIMAC project has been developed. It is based on a gaseous micro-TPC matrix, filled with $\rm ^3He$, $\rm CF_4$ and/or $\rm C_4H_{10}$. The first results on low energy nuclear recoils ($\rm ^1H$ and $\rm ^{19}F $) obtained with a low mono-energetic neutron field are presented. The discovery potential of this search strategy is discussed and illustrated by a realistic case accessible to MIMAC.}

\FullConference{Identification of Dark Matter 2010-IDM2010\\
		July 26-30, 2010\\
		Montpellier France}

\begin{document}

\section{Introduction}

Directional detection of Dark Matter is based on the fact that the solar system moves with respect to the center of our galaxy with a mean velocity of roughly 220 km/s \cite{spergel}. Taking into account the hypothesis of the existence of a galactic halo of DM formed by WIMPs (Weakly Interacting Particles) with a negligible rotation velocity, we can expect a privileged direction for the nuclear recoils in our detector, coming out from elastic collision with those WIMPs.

The MIMAC (MIcro-tpc MAtrix of Chambers) detector project tries to get these elusive events by a double detection: ionization and track, at low gas pressure with low mass target nuclei (H, 19F, 3He). In order to have a significant cross section we explore the axial, spin dependant, interaction on odd nuclei. The very weak correlation between the neutralino-nucleon scalar cross section and the axial one, as it was shown in 
\cite{PLB}, makes this research, at the same time, complementary to the massive target experiments.

\section{MIMAC prototype}

The MIMAC prototype consists of one of the chamber of the matrix allowing to show the ionization and track measurement performances needed to achieve the directional detection strategy.
The primary electron-ion pairs produced by a nuclear recoil in one chamber of the matrix are detected by driving the electrons to the grid of a bulk micromegas\cite{bulk} and producing the avalanche in a very thin gap (128 or 256$\mu$m).

\begin{figure}[h!]
\begin{center}
\includegraphics[scale=0.65]{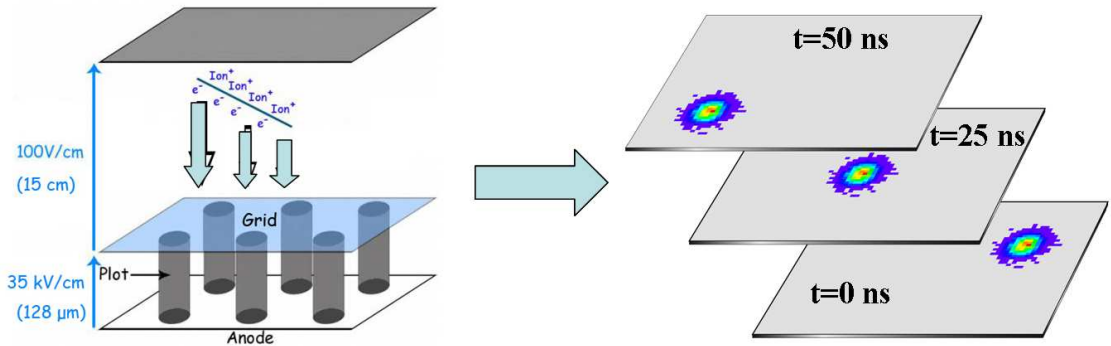}
\caption{Track reconstruction in MIMAC. The anode is read-out every 25 ns and the 3D track is recontructed,
 knowing the drift velocity of primary electrons,  from the consecutive number of images, defining the event, from the anode.}
\label{recon}
\end{center}
\end{figure}

As pictured on figure  \ref{recon}, the electrons move towards the grid in the drift space and are projected on the anode thus allowing to get 
information on X and Y coordinates.
To access the X and Y dimensions with a 100 $\mu$m spatial resolution, a bulk micromegas  with a 4 by 4 cm$^²$ active area, segmented in pixels with a pitch of 350 $\mu$m is used as 2D readout.
 In order to reconstruct the third dimension Z of the recoil, the LPSC developed a self-triggered electronics able to perform the anode sampling at a frequency of 40 MHz.
This includes a dedicated 16 channels ASIC \cite{richer} associated to a DAQ \cite{bourrion}.

In order to get the total recoil energy we need to know the ionization quenching factor (IQF) of the nuclear recoil in the gas used. We have developped at the LPSC a dedicated experimental facility to measure such IQF. A precise assessment of the available ionization energy has been performed by Santos {\it et al.} \cite{santosQuenching} in He + 5$\% \rm C_4H_{10}$ mixture within the dark matter energy range (between 1 and 50 keV) by a  measurement of the IQF.
For a given energy, an electron track in a low pressure micro-TPC is an order of magnitude longer than a recoil one. It opens the possibility to discriminate electrons from nuclei recoils by using both energy and track length informations, as it was shown in \cite{grignonMPGD}. The 3D tracks are obtained from consecutive read-outs of the anode, every 25 ns, defining the event. To get the length and the orientation of the track, an independent mesurement of the drift velocity is needed. Preliminary measurements, with different gases in the MIMAC prototype, of the drift velocity have been performed and they fit well with Magboltz simulations 
\cite{Magb}. 

\section{First experimental results}

The first result concerning the ability to detect tracks with the prototype was performed with a $\rm ^{55}Fe$ X-ray source in order to reconstruct the  5.9 keV electron tracks produced by photoelectric effect in the active volume.

\begin{figure}[h!]
\begin{center}
\includegraphics[scale=0.6]{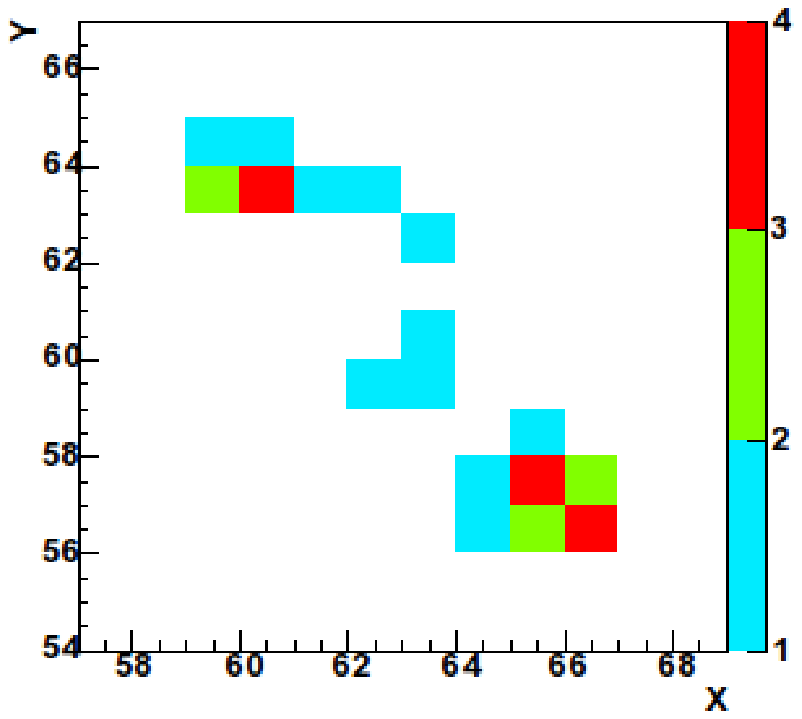}
\includegraphics[scale=0.3]{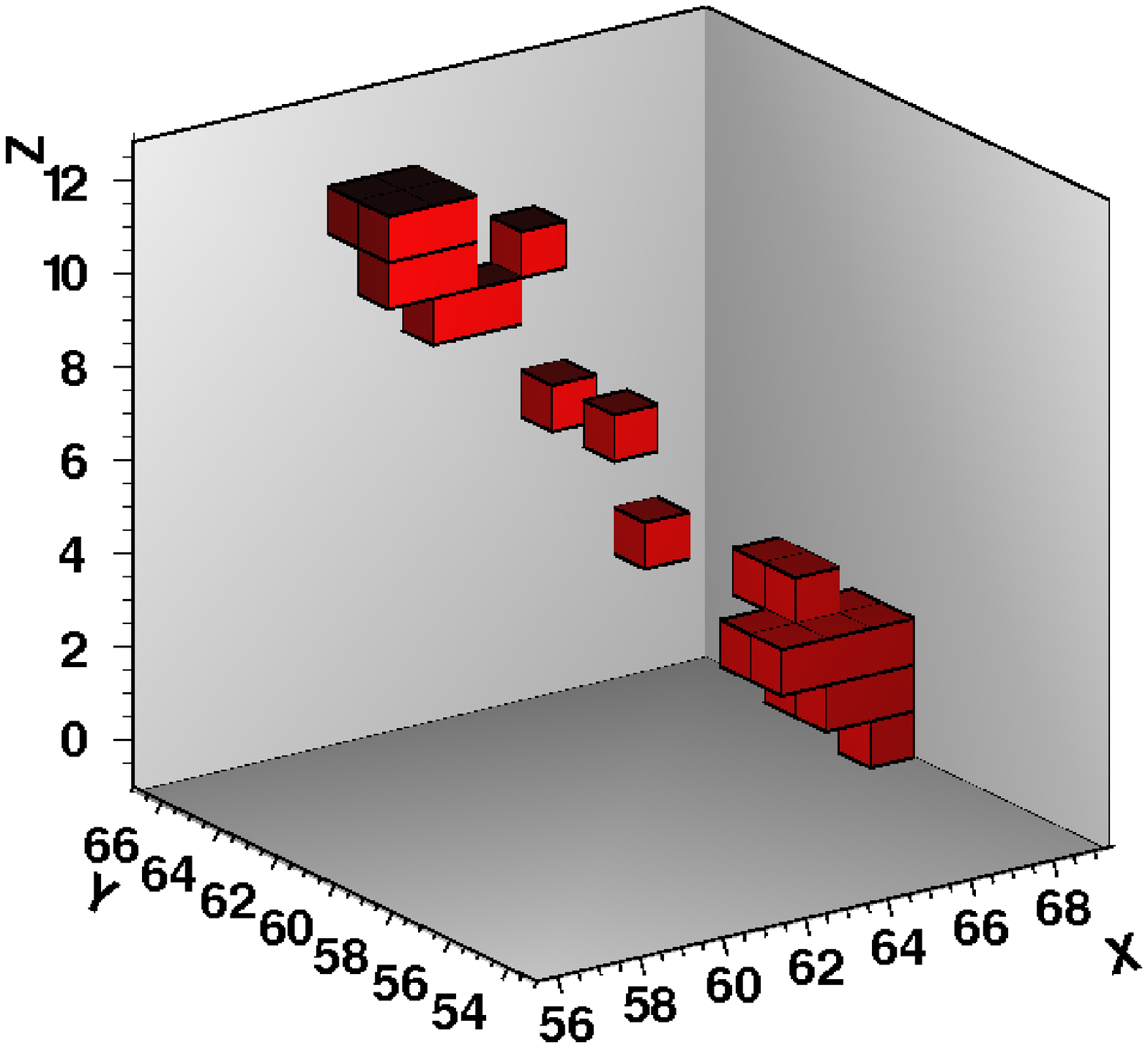}
\includegraphics[scale=0.6]{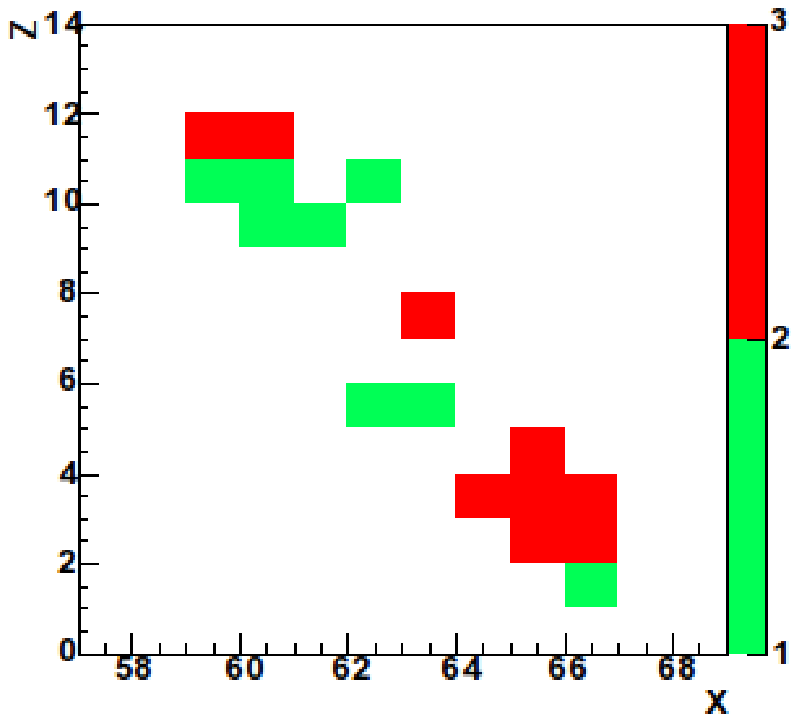}
 \caption{A 5.9 keV electron track in 350 mbar 95\%~$\rm ^4He+C_4H_{10}$.  The left panel represents the 2D projection of the recoil seen by the anode, the center panel represents a 3D view of the track after using the reconstruction algorithm and the right panel represents a projection of the 3D track on the XZ plane}
 \label{electronTrack}
\end{center}
\end{figure}

Figure \ref{electronTrack} presents a typical electron track seen by the anode (X,Y) (left panel), its projection on the XZ plane (right panel) and reconstructed in 3D (center panel).
This result shows the MIMAC capability to reconstruct the track of low energy electrons which are the  typical background in dark matter experiments.

\begin{figure}[h!]
\begin{center}
\includegraphics[scale=0.3]{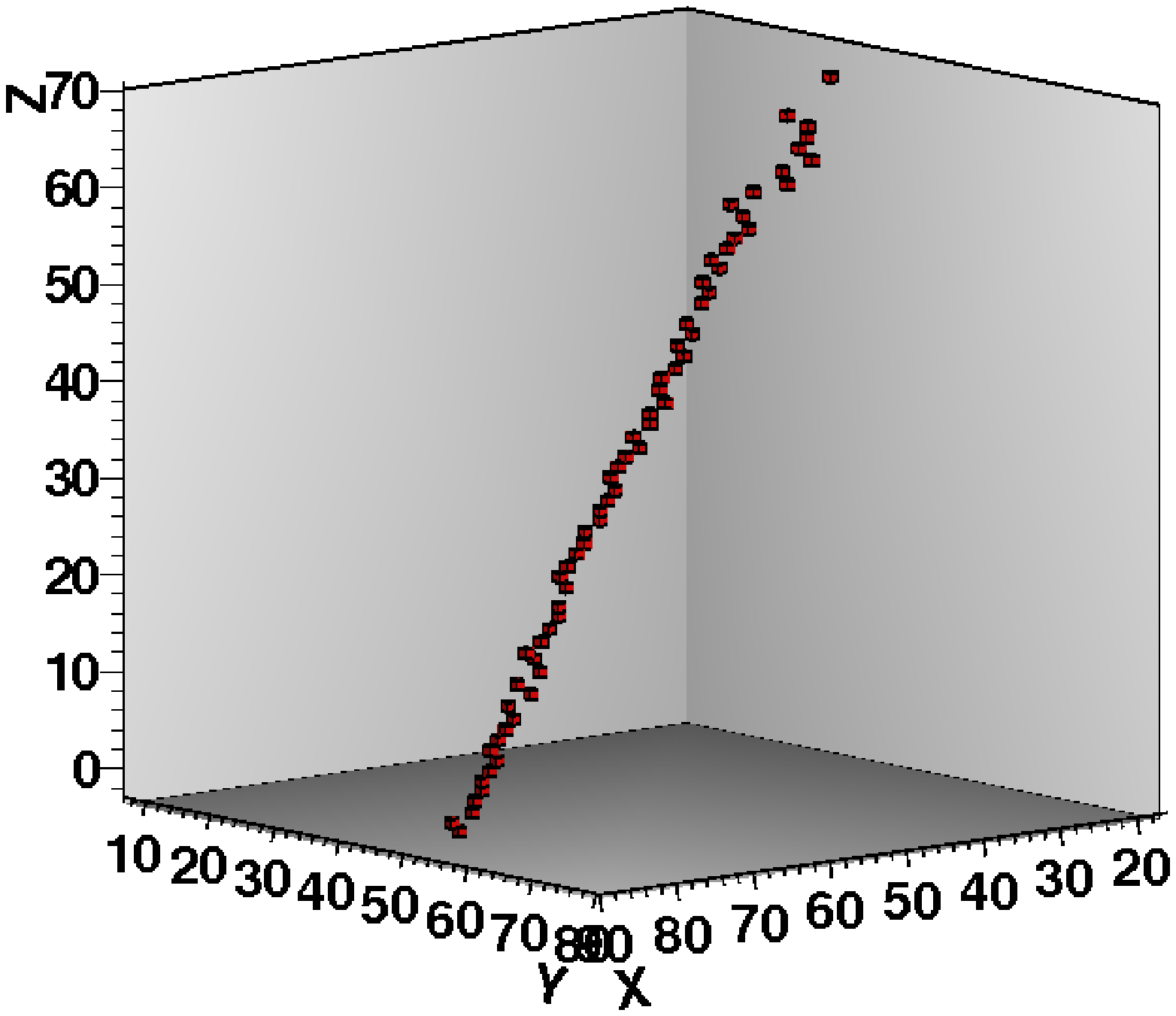}
\includegraphics[scale=0.3]{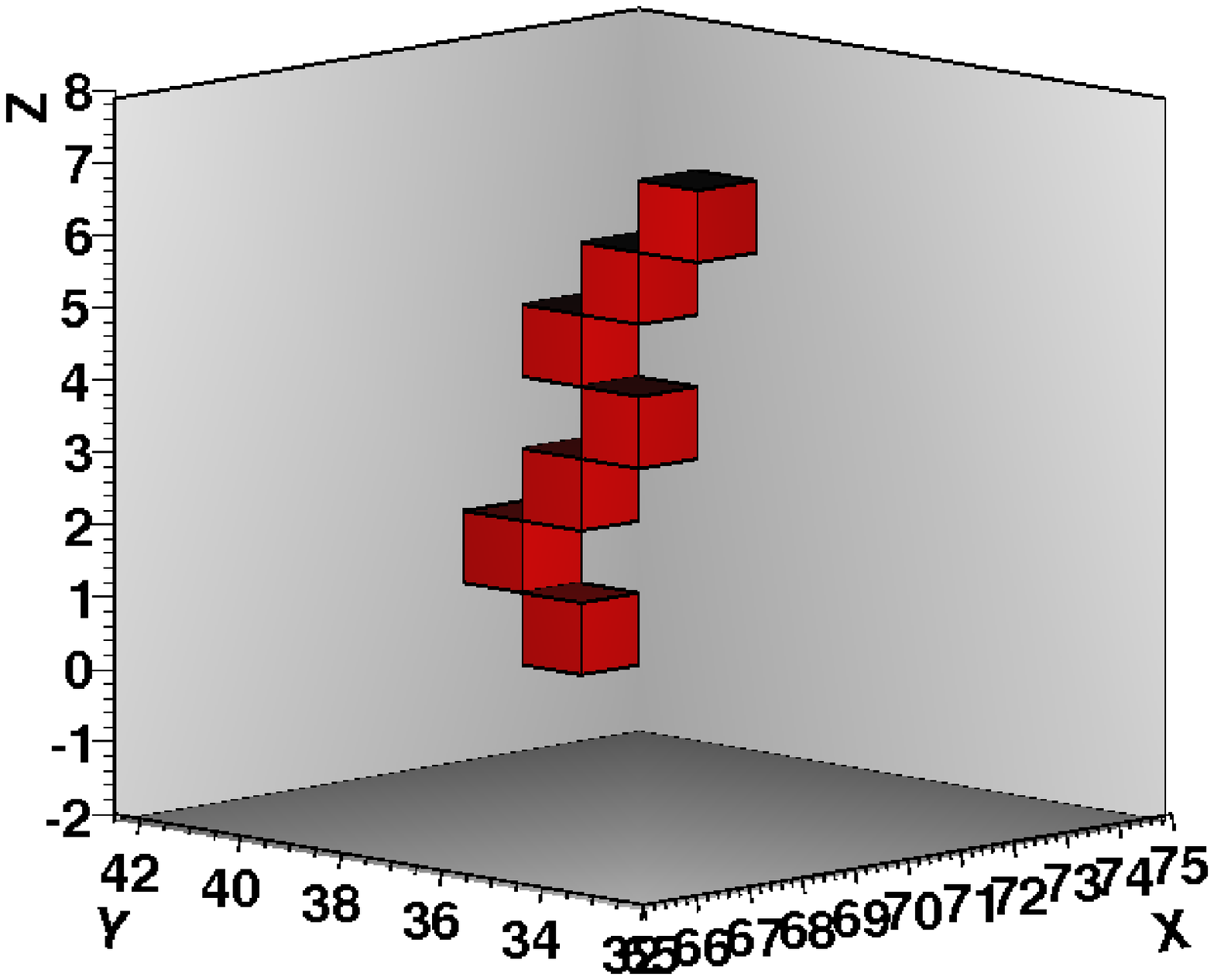}
\includegraphics[scale=0.3]{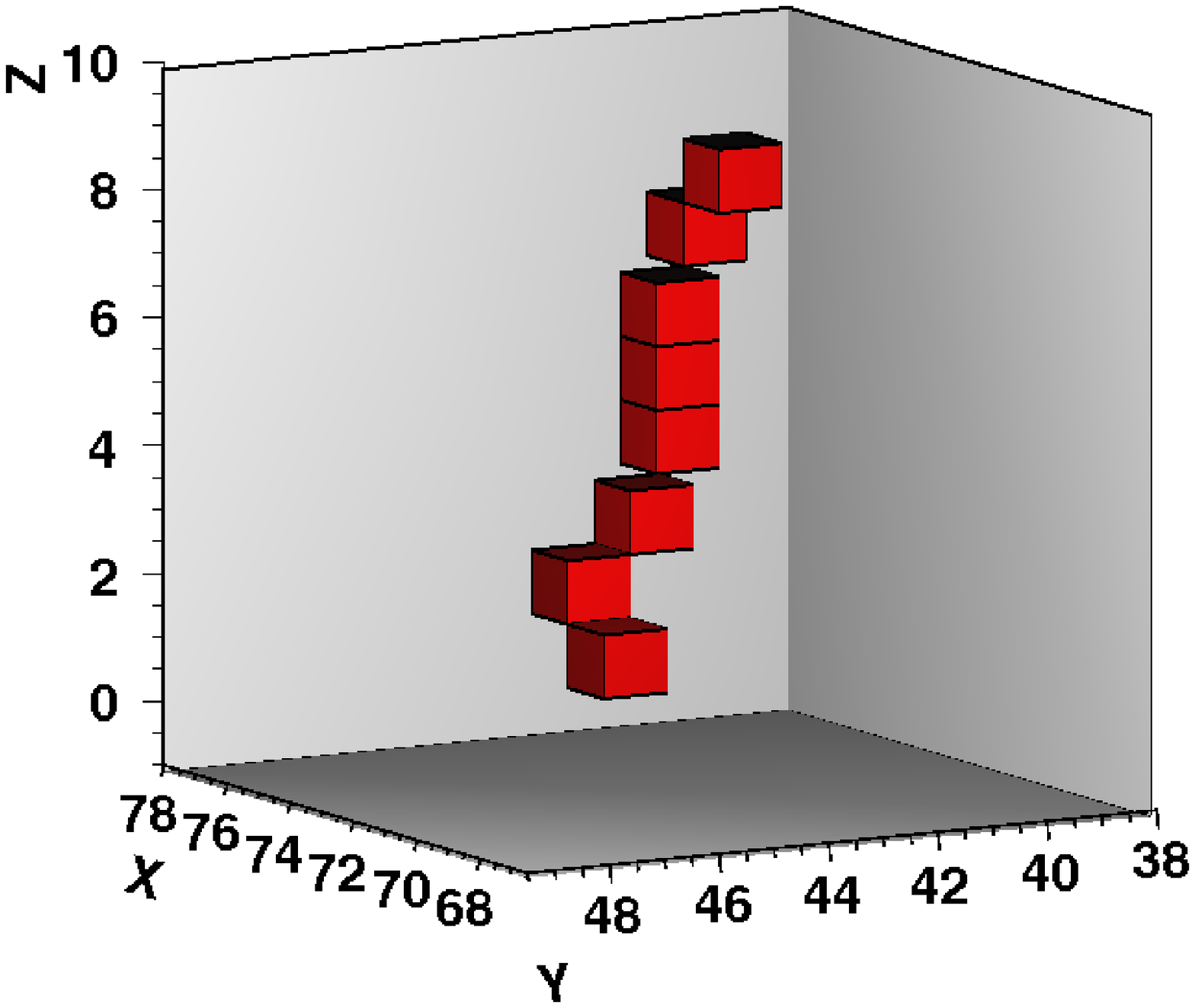}
\caption{From left to right: recoil of 5.5 MeV He nuclei  in 350 mbar $\rm ^4He+ 5\% C_4H_{10}$,  8 keV hydrogen nucleus in 350 mbar $\rm ^4He+ 5\% C_4H_{10}$ and  fluorine nucleus leaving 50 keV in ionization in 55 mbar 70\% $\rm CF_4$ + 30\% $\rm CHF_3$}
\label{nucleiTracks}
\end{center}
\end{figure}

On fig \ref{nucleiTracks} (left panel), a 3D track reconstruction is achieved for high energy (5.5 MeV) alpha particles issued from the natural radioactivity ($\rm ^{222}Rn$) present in the chamber.
However, the final validation concerning the possibility for MIMAC to get directional detection had to be done with neutrons giving nuclear recoils in the range of a few keV. In order to have monoenergetic neutron fields, in the range of a few tens of keV, we perform an experiment at the AMANDE facility (IRSN- Cadarache) allowing to select the energy of the neutrons by the angle with respect to a proton beam producing a neutron resonance on a LiF target.

On fig \ref{nucleiTracks} (center and right panel), 3D tracks of nuclear recoils following elastic scattering of mono-energetic neutrons are represented.
On the center panel, a 8 keV proton recoil leaving a track of 2.4 mm long in 350 mbar $\rm ^4He+ 5\% C_4H_{10}$ is represented. This event is a typical kind of signal that MIMAC will expect for dark matter search.
The right panel presents a 50 keV (in ionization) fluorine recoil of 3 mm long obtained in a 55 mbar mixture of 70\% $\rm CF_4$ + 30\% $\rm CHF_3$. 
In addition, the electron-recoil discrimination, a very important point for dark matter detection, showing the ability to separate  gamma background from nuclear recoils, has been presented in pure Isobutane or $\rm ^4He+ 5\% C_4H_{10}$ mixture in ref. \cite{grignonMPGD}.

\section{Directional detection of Dark Matter and its discovery potential}\label{sec:principle}

Directional detection of non-baryonic Dark Matter is a promising search strategy for discriminating  WIMP events from background offering a complementarity 
%an alternative 
to massive detectors \cite{xenon,edelweiss-armengaud}. The directionality gives a solid and unambiguous signature of the correlation between the galactic halo and a signal in our detector. 
This is achieved by searching for a correlation of the WIMP signal with  the solar motion around the galactic center, observed as a direction dependence of the WIMP flux  coming roughly from  the direction of the Cygnus constellation \cite{spergel}. 
This is generally referred to as directional  detection of Dark Matter and several projects  are being developed for 
this goal \cite{MIMAC,Drift,mit,newage,white}.\\

\begin{figure}[h!]
\begin{center}
\includegraphics[scale=0.17]{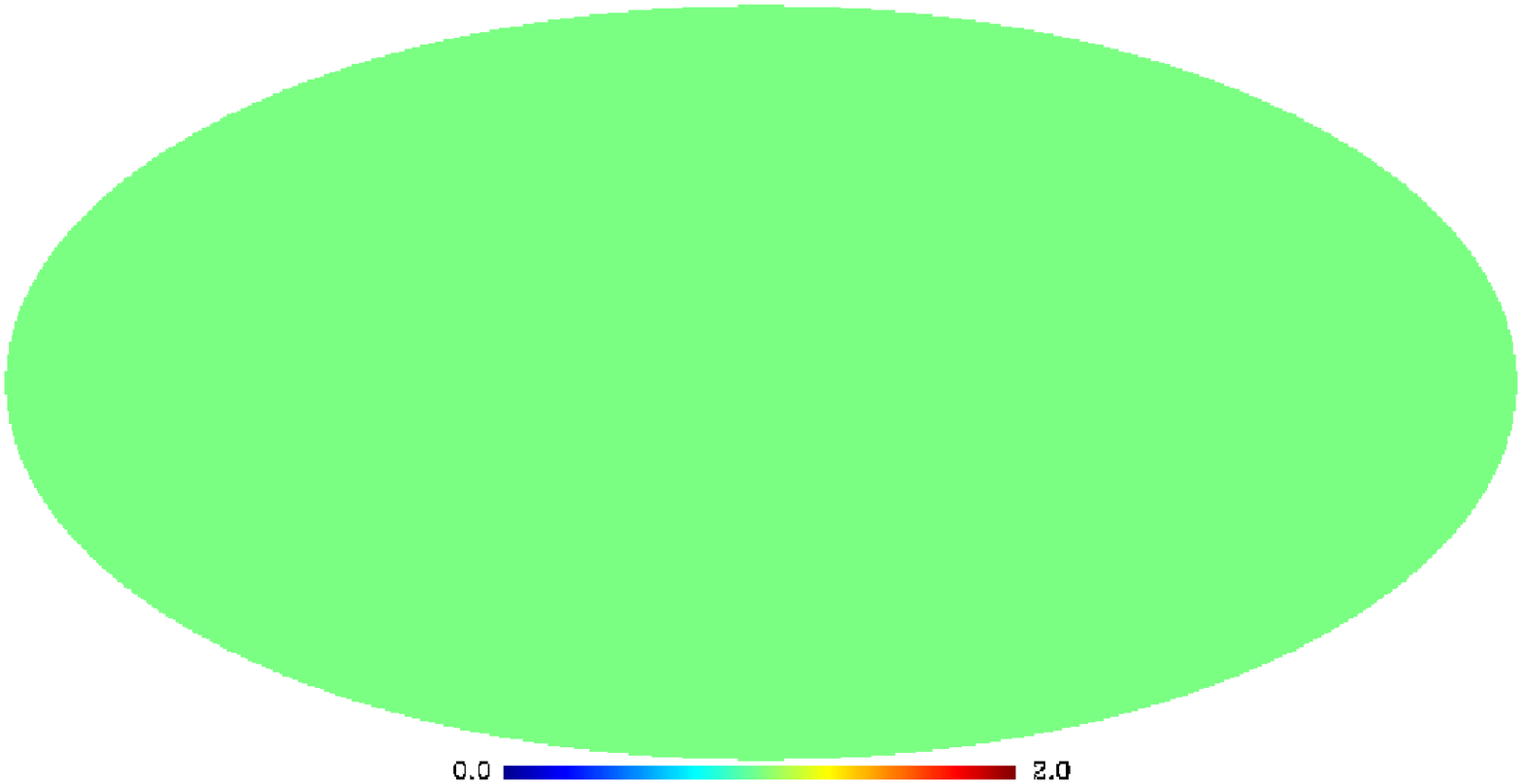}
\includegraphics[scale=0.19,angle=90]{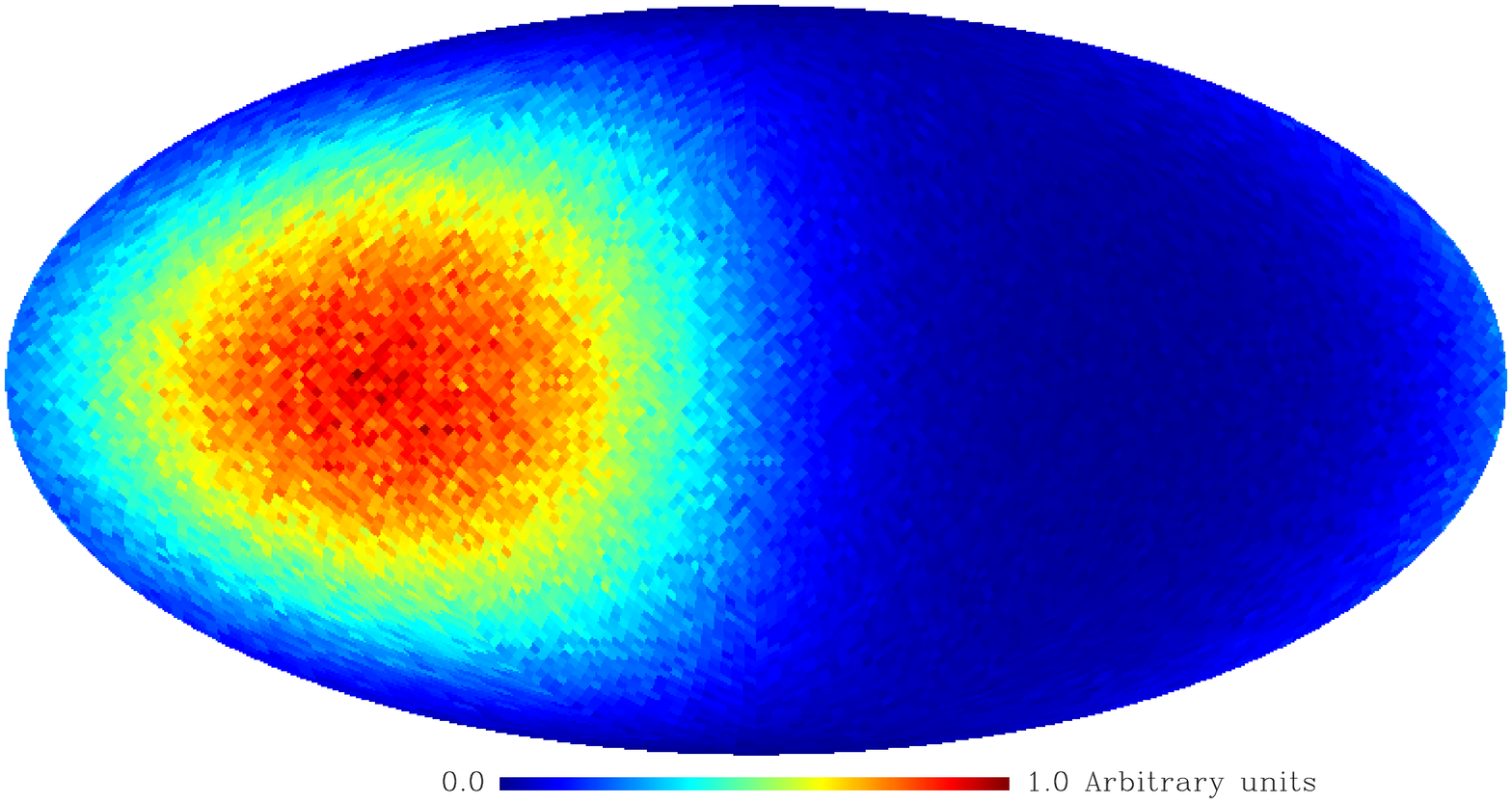}
\includegraphics[scale=0.19,angle=90]{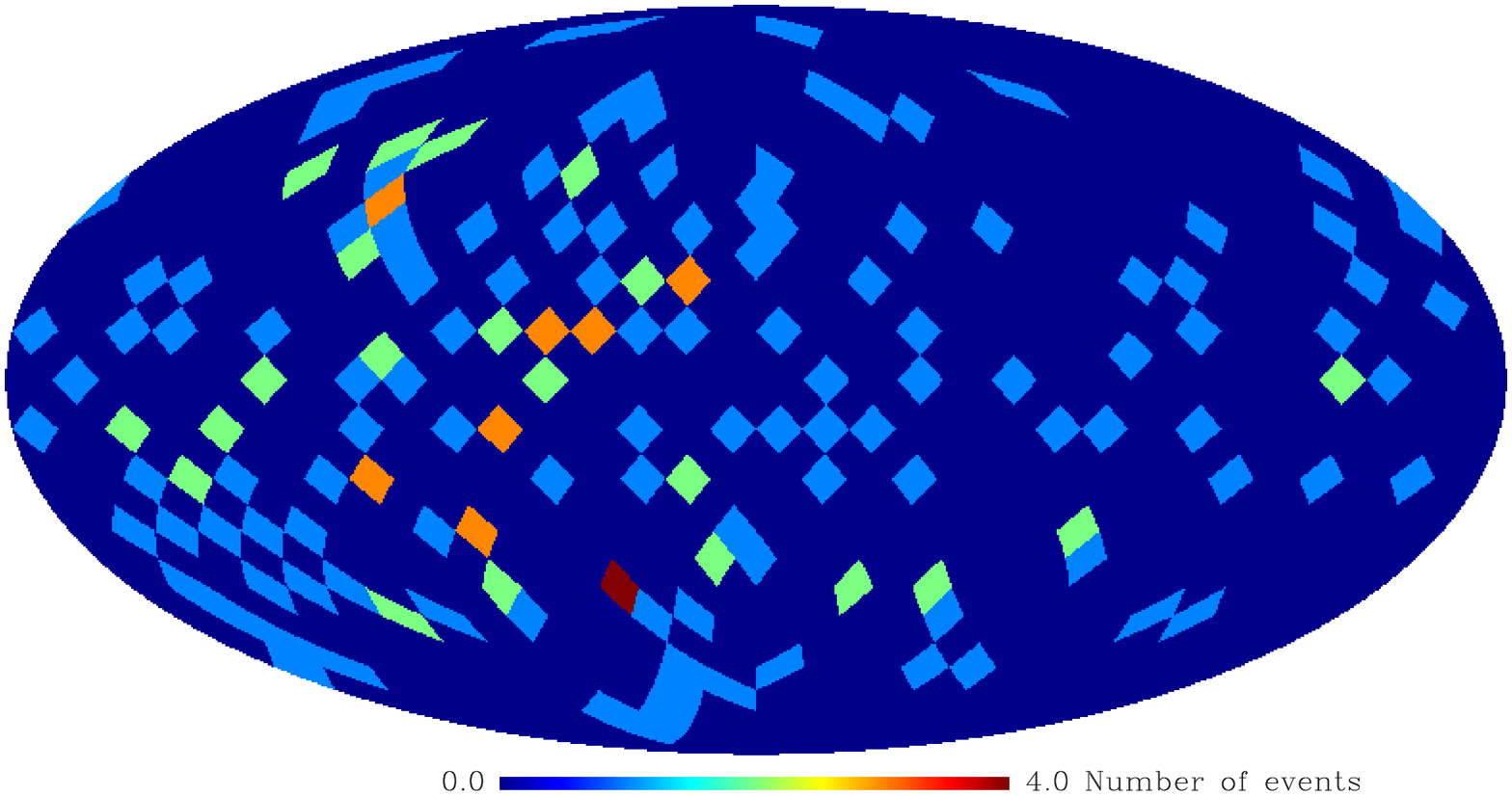}
\caption{From left to right : isotropic background distribution, WIMP-induced 
recoil distribution  in the case of an isothermal spherical halo  and a typical simulated measurement :  
100 WIMP-induced recoils and 100 background events with a low angular resolution. Recoils maps are produced for a $^{19}$F target, a 
 100 GeV.c$^{-2}$ WIMP  and considering recoil energies in the range  5 keV $\leq E_R \leq$ 50 keV. Figures from \cite{billarddisco}.}  
\label{fig:DistribRecul}
\end{center}
\end{figure}

The main asset of directional detection is the fact that the WIMP angular distribution is pointing towards the Cygnus constellation while  
  the background one is isotropic (fig \ref{fig:DistribRecul}).
The right panel of figure \ref{fig:DistribRecul}  presents a typical recoil distribution observed by a directional detector : $100$ WIMP-induced events and $100$ background events generated isotropically.  
For an elastic axial cross-section on nucleon $\rm \sigma_{n} = 1.5 \times 10^{-3} \ pb$ and a $\rm 100 \ GeV.c^{-2}$ WIMP mass, this corresponds to an exposure of $\rm \sim 7\times 10^3  \ kg.day$ in  $\rm ^{3}He$ and $\rm \sim 1.6 \times 10^3 \ kg.day$  in CF$_4$, on their equivalent energy ranges as discussed in ref. \cite{billarddisco}.
  Low resolution maps are used in this case ($N_{\rm pixels} = 768$) which is sufficient  for the low  angular resolution, $\sim 15^\circ$ (FWHM), expected for this type of detector. In this case, 3D read-out and sense recognition are considered, while background rejection  is based on electron/recoil discrimination by track length and energy  selection \cite{grignonMPGD}.
It is not straightforward to conclude from the recoil map of figure \ref{fig:DistribRecul} (right) that it does contain a fraction of WIMP events pointing towards the direction of the solar motion.

To extract information from this example of a measured map, 
%such as  the main direction of the incoming events in galactic coordinates ($\ell, b$) and the number of WIMP events contained in the map, 
a likelihood analysis has been developed.
The likelihood value is estimated using a binned map of the overall sky with  Poissonian statistics,  as shown in Billard {\it et al.} \cite{billarddisco}.
This is a four parameter likelihood analysis with $m_\chi$,  $\lambda = S/(B+S)$ the  WIMP fraction ($B$ is the  background spatial distribution taken as isotropic and $S$ is the WIMP-induced recoil distribution) and the coordinates ($\ell$, $b$) referring to the maximum of the WIMP event angular distribution.

\begin{figure}[h]
\begin{center}
\includegraphics[scale=0.45]{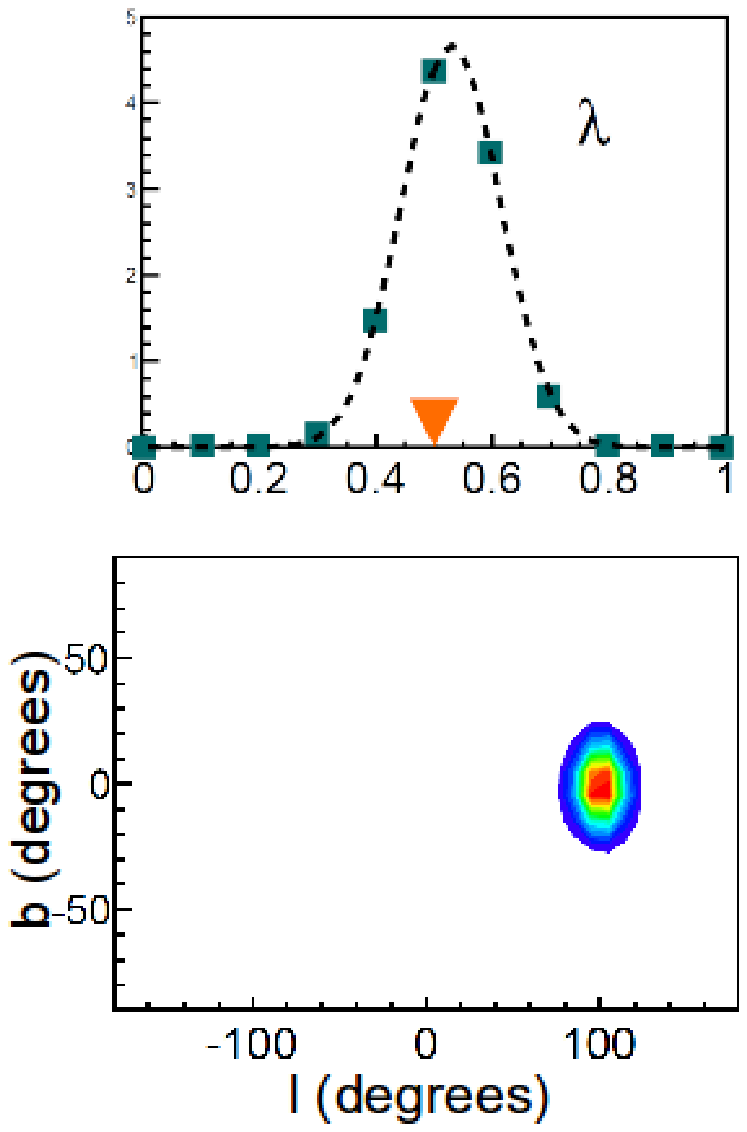}
\includegraphics[scale=0.30]{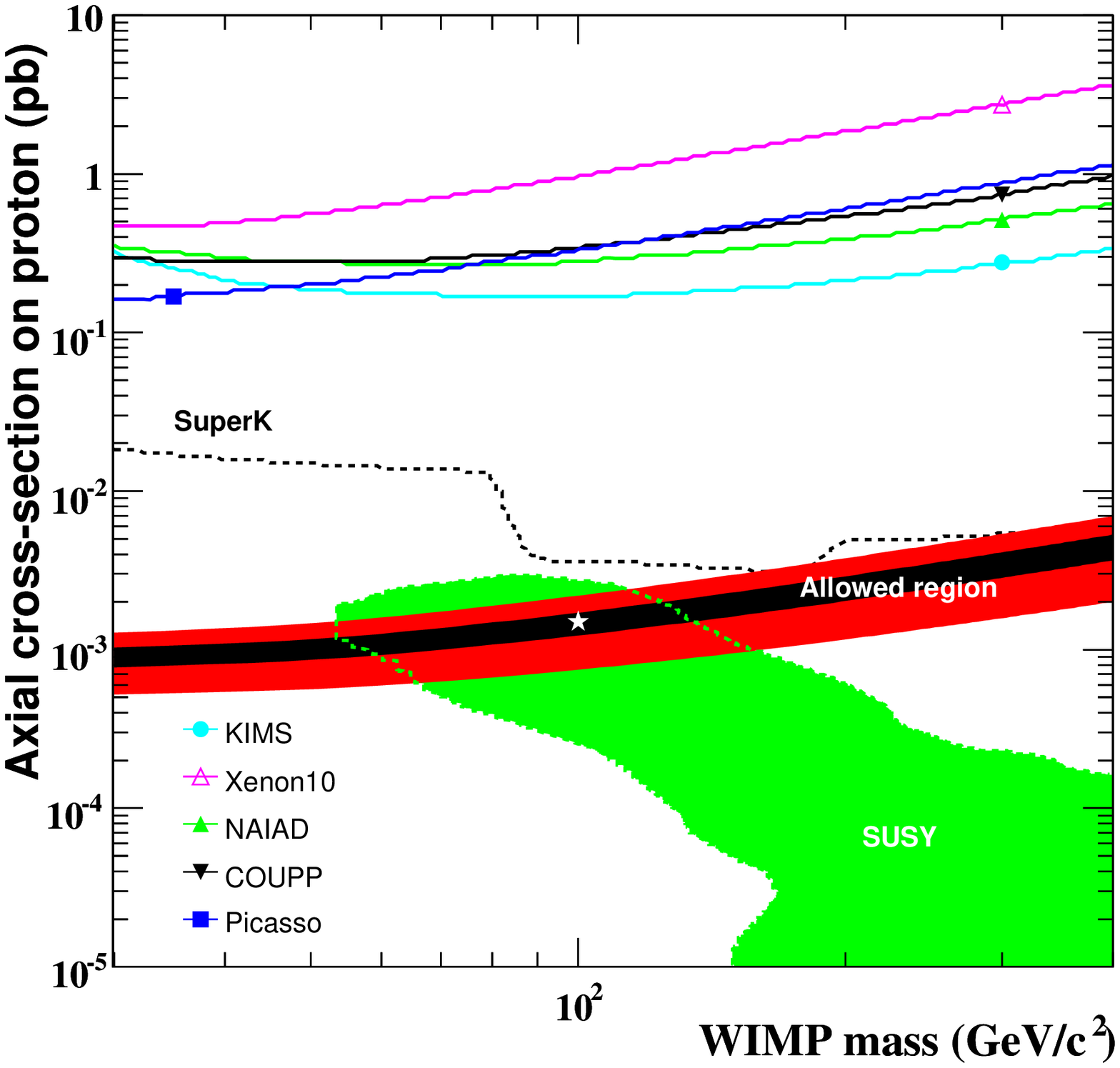}
\caption{On the left, marginalized probability density functions 
 of $\lambda$ (top), $\ell, b$ (bottom) after the likelihood analysis of the simulated recoil map shown before. On the right, allowed regions 
  presented in the plane of the spin dependent cross-section on proton (pb) as a function of the WIMP mass (GeV/c$^2$).
 Input value for the simulation is shown with a star.}
\label{fig:DiscAndExcl}
\end{center}
\end{figure}

The result of this  map-based likelihood method is that the main recoil direction is recovered and it is  pointing towards ($ \ell = 95^{\circ} \pm 10^{\circ}, b = -6^{\circ} \pm 10^{\circ}$) at $68 \  \%$ CL, corresponding to a non-ambiguous detection of particles from the galactic halo. This is indeed the discovery  proof of this detection strategy (left panel of fig. \ref{fig:DiscAndExcl}) \cite{billarddisco}.
Furthermore, the method allows to constrain the WIMP fraction in the observed recoil map leading to a constraint in the $(\sigma_n, m_\chi)$ plane (right panel of fig. \ref{fig:DiscAndExcl}).  
As emphasized in ref. \cite{billarddisco}, a directional detector could allow for a high 
significance discovery of galactic Dark Matter even with a sizeable background contamination.
For very low exposures, competitive exclusion limits may also be imposed \cite{billardexclu}.

\section{Conclusions}

Directional detection is a promising search strategy to discover galactic dark matter.
The MIMAC detector provides the energy of a recoiling nucleus and the reconstruction of its 3D track. 
The first 3D tracks  observed with the MIMAC prototype were shown: 5.9 keV electrons (typical background) and low energy proton and fluorine recoils (typical signal).
The next step will be to build a demonstrator of 1 m$^3$ to show that the large micro-tpc matrix for directional detection of dark matter search is accessible.

\section{Acknowledgements}

The MIMAC collaboration acknowledges the ANR-07-BLANC-0255-03 funding.

\end{document}